\documentclass[utf8]{jetp} 
\twocolumn 
\usepackage{amsmath}

\begin{document}

\title{Black Hole Shadows Modelling in Extended gravity: Rotation Accounting and coupled effects}

\thanks{Black Hole Shadows Modelling...}%

\setaffiliation1{Sternberg Astronomical Institute, Lomonosov Moscow State University, Universitetskii Prospekt, 13, Moscow, 119234, Russia}

\setaffiliation2{Department of Quantum Theory and High Energy Physics, Physics Faculty, Lomonosov Moscow State University, Vorobievi Gori, 1/2, Moscow, 119234, Russia}

\setauthor{S.}{Alexeyev}{12}
\setauthor{O.}{Zenin}{2}
\setauthor{A.}{Baiderin}{2}

\date{October 16, 2024}

\abstract{
Using an improved version of the Newman-Janis algorithm, we obtain metrics of rotating black holes for a set of extended gravity theories that extend general relativity in different ways: the Horndeski model, the bumblebee model, the Gauss-Bonnet scalar gravity, the loop quantum gravity, the conformal gravity, and $f(Q)$ (symmetric teleparallel gravity STEGR). The obtained metrics are used to model black hole shadows. It is shown that for some models the critical values of the angular momentum $a_{crit}$ emerge. The previous conclusion that the extended gravity theory corrects by itself the effects of rotation in both directions is confirmed. This appears to be important for further modelling of shadow profiles taking into account the constantly increasing accuracy of black hole images. Thus, when rotation is taken into account black hole shadow images, as the GW170817 test or the post-Newtonian formalism, can serve as a meaningful way to test and constrain extended theories of gravity.

}

\maketitle


\section{Introduction}\label{sec1}

Temporary appearance of new data from Event Horizon Telescope (EHT) with the increasing accuracy \cite{EventHorizonTelescope:2023fox} leads to the necessity to improve the accuracy of black hole shadows theoretical modelling. The first models were constructed based on spherically symmetric metrics \cite{Zakharov:2005bc}. Next step means accounting of the electrical or tidal charge \cite{Dadhich:2000am} from Reissner-Nordstrom metrics. Such tidal charge application allows to take into account the contribution of a new physics \cite{Zakharov:2014lqa}. Further it was shown that taking into account both the tidal charge and the next perturbation order terms can improve the quality of the shadow modelling to achieve the best fit of EHT the results (including an eye to the future) \cite{Alexeyev:2018ofs,Alexeyev:2022mqb}. 

From the beginning it was shown that both black holes, i.e. M87 and Sgr A, are spinning \cite{EventHorizonTelescope:2022apq,EventHorizonTelescope:2019pgp}. Their angular velocity was measured not so long ago \cite{Cui:2023uyb}. So for future investigations better to use the metric that is analogous to Kerr-Newman one \cite{Prokopov:2020ewm}. The idea to measure the parameters of spinning black hole was discussed earlier in application to RadioAstron \cite{Zakharov:2005newastr}. Here it is necessary to note that in each new gravity model the first obtained solution appears to spherically-symmetric as most simple from mathematical view. The direct solution of Einstein equations in axially-symmetric case is not very simple \cite{Chandrasekhar:1985kt}. That is why Newman and Janis suggested an alternative method to generate rotation \cite{Newman:1965tw}. Last years the interest to this method increased because it was re-established in the form of a classical algorithm \cite{Azreg-Ainou:2014aqa}. Thanks to using such an algorithm new relations between the tidal charge and the moment of rotation in the frames of GR \cite{Karas:2023uac}, for bumblebee model  \cite{Capozziello:2023tbo} and for non-local gravity models \cite{c02}. Further the effect of rotation increasing and decreasing was found. In addition, shadow profile depends upon the other theory parameters. So if after increasing the EHT accuracy it would revealed that the shadow profile does not correspond to its Kerr value this will mean the possibility to measure the contribution from the new physics. Therefore we study how the characteristics of the shadow would change when rotation was taken into account in various modern extended theories of gravity. We proposed a set of theories extending GR in different ways including the Horndeski model (the most general case of scalar-tensor gravity with second-order field equations), bumblebee (gravity with a vector field), scalar Gauss-Bonnet gravity, loop quantum gravity, conformal gravity and $f(Q)$ --- a special case of symmetric teleparallel gravity STEGR. We have previously modelled the  shadows profiles for non-rotating cases of these theories \cite{Prokopov:2021lat}. Preliminary results of rotating metrics and generation and shadow profiles modelling for the Horndeski and bumblebee models were presented earlier \cite{c03}, but we consider it is necessary to present their full versions for a more visual comparison of the modelling results.  

It seems to be necessary to point out an important feature of the Newman-Janis algorithm. Note that this algorithm implements only a simple one-parametric rotation. In general the symmetry group of the axially symmetric solution could be wider. This means that after using this algorithm the resulting axially symmetric solution will represent a particular case (for example, similar to the Kerr-like solution in $D>4$ with only one momenta). Since our goal is to estimate changes in the black hole shadow profile when the BH rotation is taken into account. We believe that the usage of the Newman-Janis algorithm to generate a rotating solution (even if degenerate) is an important step in this direction. Further, we note that for the Horndeski theory we consider a particular spherically symmetric solution. At the same time, the number of BH metrics being the solutions for the Horndeski model, is much wider. Unfortunately, neither we nor other researchers have yet been able to find a general solution to this theory.

The structure of this article is as follows: Section 2 is devoted to the discussion of how to obtain axially symmetric solutions and, in fact, the generation of these solutions for the above-mentioned models, Section 3 is devoted to the results of shadow profiles modelling for each of the theories under consideration, and Section 4 contains discussion and our conclusions.

\section{Obtaining solutions with rotation}

Following \cite{Azreg-Ainou:2014aqa} we consider spherically symmetric metrics parameterized by metric functions $G(r)$, $F(r)$ и $H(r)$ in the form: 
\begin{equation}\label{metric} 
ds^2 =  - G(r) dt^2 + \frac{1}{F(r)} dr^2 + H(r)d\Omega^2.
\end{equation}
This redefined parametrization allows to consider the widest possible class of spherically symmetric solutions. Further development of the Newman-Janis algorithm reduced the generation of an axially symmetric solution to the calculation of a new metric with the following components:
\begin{eqnarray}\label{spin_metri}
g_{tt} & = & -\frac{FH + a^2\cos^2{\theta}}{(K + a^2\cos^2{\theta})^2} \Psi,\nonumber \\   
g_{t\phi} & = & -a\sin^2{\theta}\frac{K - FH}{(K + a^2\cos^2{\theta})^2}\Psi,\nonumber \\
g_{\theta \theta} & = & \Psi,\nonumber \\ 
g_{rr} & = & \frac{\Psi}{FH + a^2}, \\
g_{\phi \phi} & = & \Psi\sin^2{\theta}(1 + a^2\sin^2{\theta}\frac{2K - FH + a^2\cos^2{\theta}}{(K + a^2\cos^2{\theta})^2}). \nonumber
\end{eqnarray}
Here $K = H(r)\sqrt{F(r)/G(r)}$, and all components are defined up to a function $\Psi(r,y^2,a)$, where $y \equiv \cos{\theta}$. This additional function $\Psi(r,y^2,a)$ must follow the conditions:
\begin{eqnarray}\label{dif_eq}
&& \lim_{a\to 0} \Psi(r,y^2,a) = H(r), \nonumber \\
&& (K + a^2y^2)^2(3\Psi_r\Psi_{y^2} -2\Psi\Psi_{r,y^2}) = 3a^2K_r\Psi^2,\nonumber\\
&& \Psi \left[ K^2_r + K(2 - K_{rr}) -a^2y^2(2 + K_{rr})\right] + \nonumber \\ & + & (K + a^2y^2)[(4y^2\Psi_{y^2} - K_r\Psi_r] = 0.
\end{eqnarray}
First condition from Eqs. (\ref{dif_eq}) means that when $a \to 0$ the solution transforms into the metric of a non-rotating black hole with 2 representations: ($\Psi_n$ и $\Psi_c$). The representations are related via a conformal transformation, so the initial metric can be represented as:
\begin{eqnarray}
    ds_{c}^{2} = \Psi_c/\Psi_nds^2_n.
\end{eqnarray}
So the solutions of Eq. (\ref{dif_eq}) are looked for in the form:
\begin{eqnarray}
\Psi_c & = & H(r)\exp{[a^2f(r,a^2y^2,a)]} \approx \nonumber \\
& \approx & H(r) + a^2X(y^2,r) + o(a^2),
\end{eqnarray}
where Taylor expansions used, 
$A_r = \partial A/\partial r$ and: 
\begin{eqnarray}\label{new_dif_eq}
&& KH_rK_r + HK_r^2 + HK(K_{rr} -2) = 0, \nonumber \\ 
&& X(y^2,r) = \frac{H^2(8K - K_{r}^2)y^2}{K^2(8H - H_rK_r)}, \nonumber \\
&& K_r(8K - K_r^2)K_{rrr} + K_r^2(K_{rr} - 2)^2 - \nonumber \\ 
& - & 4 KK_{rr}(K_{rr} + 4) + 48K = 0.
\end{eqnarray}

\subsection{Horndeski theory}

Previously We considered the spherically symmetric BH metric being one of the particular solutions of the Horndeski theory \cite{Babichev:2017guv,Prokopov:2021lat,c03}:
\begin{eqnarray}\label{horn} 
ds^2 =  & - & (1 - \frac{2M}{r} - \frac{8\alpha_{5}\eta}{5r^3})dt^2 + \nonumber \\ & + & \frac{1}{1 - \frac{2M}{r} - \frac{8\alpha_{5}\eta}{5r^3}}dr^2 + r^2d\Omega^2 ,
\end{eqnarray}
where $\alpha_5$ and $\eta$ are theory parameters. After the Newman-Janis algorithm application (in the form described in the previous paragraph) we obtain a rotating metric in the form ($\rho^2=r^2+a^2\cos^2\theta$, all the other metric components vanish):
\begin{eqnarray}\label{spin_metri_horn1}
g_{tt} & = & - \left( 1 - \frac{2Mr}{\rho^2} - \frac{8 \alpha_5 \eta}{5r}\right) ,\nonumber \\ 
g_{t\phi} & = & - \frac{2a\sin^2\theta}{5r\rho^2} \left( 4\alpha_5 \eta + 9 Mr^2 \right),\nonumber \\
g_{rr} & = & \rho^2 \left( - \frac{8\alpha_5\eta}{5r} + a^2 - 2Mr + r^2  \right)^{-1}, \nonumber \\
g_{\theta \theta} & = & \rho^2 ,\nonumber \\
g_{\phi\phi} & = & \frac{\sin^2\theta}{\rho^2} \biggl(r^4 + 2 a r^2 \cos^2\theta + a^4 \cos^4\theta + \nonumber \\ & + & \frac{8a^2\alpha_5\eta\sin^2\theta}{5r} + 2aMr\sin^2\theta + \nonumber \\ & + & a^2 r^2 \sin^2\theta + a^4 \cos^2\theta\sin^2\theta \biggr),
\end{eqnarray}

\subsection{Bumblebee model}

The action for bumblebee field $B_{\mu}$ is \cite{Casana:2017jkc}:
\begin{eqnarray}\label{b1}
S_B & = & \int d^{4}x\mathcal{L}_B=\int d^{4}x(\mathcal{L}_g + \mathcal{L}_{gB} +
\mathcal{L}_K + \nonumber \\ & + & \mathcal{L}_V + \mathcal{L}_M),
\end{eqnarray}
where $\mathcal{L}_g$ is GR action, $\mathcal{L}_{gB}$ represents the coupling between gravity and bumblebee field, $\mathcal{L}_K$ are the kinetic term of bumblebee field and other self-action terms, $\mathcal{L}_V$ is the potential from spontaneous Lorenz symmetry breaking, $\mathcal{L}_M$ is matter and its interaction with bumblebee field. When the torsion and the cosmological constant vanish:
\begin{eqnarray}\label{b2}
\mathcal{L}_B & = & \frac{e}{2\kappa}R+\frac{e}{2\kappa}\xi B^{\mu}B^{\nu}R_{\mu\nu} - \nonumber \\ & - & \frac{1}{4}eB_{\mu\nu}B^{\mu\nu} - eV(B^{\mu})+\mathcal{L}_M ,
\end{eqnarray}
where $e=\sqrt{-g}$ is a constant of non-minimal coupling between gravity and bumblebee field. 

We start from the metric in the form \cite{Casana:2017jkc,Prokopov:2021lat,c03}:
\begin{eqnarray}\label{metric_bb} 
ds^2 =  &-& (1 - \frac{2M}{r}) dt^2 + \frac{1 + l}{1 - \frac{2M}{r}}dr^2 + r^2d\Omega^2 ,
\end{eqnarray}
where $l$ is bumblebee parameter. After applying the Newman-Janis algorithm the solution looks like (as before all the other metric components vanish and the values $A$-$K$ are used only for the discussed case):
\begin{eqnarray}\label{spin_metri_bb}
g_{tt} & = & \frac{r^{-1+\sqrt{1+l}} A B }{\sqrt{1+l} C D}, \nonumber \\ 
g_{t\phi} & = & -\frac{ar^{-l+\sqrt{1+l}} E B \sin^2{\theta}}{(1+l)C D}, \nonumber \\ 
g_{rr} & = & \frac{(1+l)r^{-l+\sqrt{1+l}} B}{C G},\nonumber \\
g_{\theta \theta} & = & r^{1+\sqrt{1+l}}+\frac{a^2(-4+8\sqrt{1+l})r^{-l+\sqrt{1+l}}\cos^2{\theta}}{8-2(1+\sqrt{1+l})}, \nonumber \\
g_{\phi \phi} & = &\frac{r^{-l+\sqrt{1+l}}\sin^2{\theta}(B+5a^2\cos^2{\theta})}{(1+l)C D} \times \nonumber \\ & \times & \Bigl( D(1+l)-Ka^2\cos^2{\theta}\Bigr),  
\end{eqnarray}
where 
\begin{eqnarray*}
A & = & (2Mr^{1+l}-r^{1+\sqrt{1+l}}-a^2\cos^2{\theta}-a^2l\cos^2{\theta}), \\
B & = & -3r^2 + \sqrt{1+l} r^2 - 3a^2\cos^2{\theta} - 4a^2\sqrt{1+l}\cos^2{\theta}, \\
C & = & -3+\sqrt{1+l}, \qquad D = r^2+a^2\sqrt{1+l}\cos^2{\theta}, \\
E & = & -r^2-lr^2-2\sqrt{1+l}Mr^{\sqrt{1+l}}+\sqrt{1+l}r^{1+\sqrt{1+l}}, \\
G & = & a^2+a^2l-2Mr^{1+l}+r^{1-\sqrt{1+l}}, \\
F & = & -2Mr^{\sqrt{1+l}}+r^{1+\sqrt{1+l}}-a^2 l\cos^2{\theta}, \\
K & = & \sqrt{1+l}F-r-2lr^2-D.
\end{eqnarray*}

\subsection{Gauss-Bonnet scalar gravity}

Gauss-Bonnet scalar gravity as a model includes all possible second-order curvature corrections in the form \cite{Yunes:2011we,Prokopov:2021lat}:
\begin{eqnarray}\label{sgb1}
S & = & \int d^4x\sqrt{-g}\biggl[ \kappa 
 +\alpha_1 f_1 (\vartheta)R^2 + \alpha_2f_2 (\vartheta) R_{ab}R^{ab} + \nonumber \\ & + & \alpha_3 f_3 (\vartheta) R_{abcd}R^{abcd} + \alpha_4f_4 (\vartheta) R_{abcd}^* {R^*}^{abcd} - \nonumber \\ & - &  
\frac{\beta}{2}\Bigl( \nabla_a \vartheta \nabla^a \vartheta +2V (\vartheta) \Bigr) + \mathcal{L}_{mat}\biggr].
\end{eqnarray}
Here $g$ is the metrics $g_{ab}$ determinant, ($R$, $R_{ab}$, $R_{abcd}$ and $R_{abcd}^*$) are Ricci scalar, Ricci and Riemannian tensors and Riemannian tensors, $\mathcal{L}_{mat}$ is matter Lagrangian, $\vartheta$ is scalar field, ($\alpha_i, \beta$) are coupling constants, $\kappa=(16\pi G)^{-1}$. The spherically-symmetric space-time is parametrised via metric functions from Eq. (\ref{metric}) as:
\begin{eqnarray}
G(r) &=& f_s \left(1 + \frac{\xi}{3r^3f_s}\right) + o\left(\frac{1}{r^3}\right), \nonumber \\
F(r) &=& \frac{f_s}{(1 - \frac{\xi}{r^3f_s})} + o\left(\frac{1}{r^3}\right), \nonumber \\
H(r) & = & 2\frac{K}{K_r}r, 
\end{eqnarray}
where $f_s = 1 - 2M/r$. After applying of the Newman-Janis algorithm the solution looks like (as before, all other metric components vanish and the values $A$-$T$ are used only for the discussed case):
\begin{eqnarray}\label{spin_metri_gb}
g_{tt} & = & \frac{r^2(E+F\cos^2{\theta})}{AB}, \nonumber \\
g_{t\phi} & = & -\frac{aCD\sin^2{\theta}}{AB},\nonumber \\
g_{rr} & = &-\frac{AB}{r^2(E+F)}, \nonumber \\
g_{\theta \theta} & = & \frac{B}{3r^2}, \nonumber \\ 
g_{\phi \phi} & = & \frac{T}{3r^2AB},
\end{eqnarray}
where
\begin{eqnarray*}
A & = & \xi+2Mr^2-r^3, \\
B & = & 2\xi M+\xi r+3r^4+3a^2r^2\cos^2{\theta}, \\ 
C & = & 2\xi M+\xi r+3r^4, \\
D & = & A+16M^2r^2-16Mr^4+4r^5, \\
E & = & 32\xi M^3r-16\xi M^2r^2-8\xi Mr^3+4\xi r^4 + \\ & + & 48M^2r^5-48Mr^6+12r^7, \\
F & = & -3a^2\xi-6a^2Mr^2+3a^2r^3, \\ 
G & = & 16\xi M^3r^5+2\xi r^6+24M^2r^7-24Mr^8+6r^9, \\
K & = & 2\xi^2M+\xi^2r+4\xi M^2r^2+2\xi r^4+6Mr^6-3r^7, \\
Q & = & 4\xi^3 M(M+r)+\xi^2 r^2(\xi+2M^3+4M^2r +  \\ & + & 10Mr^2+5r^3)+3\xi r^6(8M^2+r^2)+9r^{10}(2M-r), \\
T & = & 1 + Q + 9a^4 r^4 A \cos^4{\theta} + 6 a^2 r^2 G \sin^2{\theta} + \\ & + &  9a^4r^4A\cos^2{\theta} \sin^2{\theta}.
\end{eqnarray*}

\subsection{Loop Quantum Gravity}

Now we consider a modified Hayward metric: a BH without a central singularity   \cite{De_Lorenzo_2015,Hu_2018,Prokopov:2021lat}, where: 
\begin{eqnarray}\label{loop} 
G(r) & = & \left(1 - \frac{2Mr^2}{r^3+2Ml^2}\right)\left(1 - \frac{M\alpha\beta}{\alpha r^3+\beta M}\right), \nonumber \\
F(r) & = & 1 - \frac{2Mr^2}{r^3+2Ml^2}, \nonumber \\
H(r) & = & r^2. 
\end{eqnarray}
Applying the modified Newman-Janis method we obtain a rotating solution for conformal gravity ($\rho^2=r^2 + a^2\cos^2\theta$):
\begin{eqnarray}\label{spin_metri_loop_1}
g_{tt} & = & -\frac{\rho^2 - A}{B}, \nonumber \\   
g_{t\phi} & = & -a\sin^2{\theta}\frac{\Tilde{H} - r^2(1 - A)}{B}, \nonumber \\
g_{\theta \theta} & = & B,\nonumber \\ 
g_{rr} & = & \frac{B}{r^2 + a^2 -A}, \nonumber \\
g_{\phi \phi} & = & \sin^2{\theta}\left (B + a^2\sin^2{\theta}\frac{2\Tilde{H} - 2r^2 +\rho^2 + r^2 A}{B}\right),
\end{eqnarray}
where 
\begin{eqnarray*}
\Tilde{H}(r) & = & r^2 \left( 1 - \frac{M\alpha\beta}{\alpha r^3+\beta M}\right)^{-1/2} , \\
A & = & 2Mr^2 \left(r^3+2Ml^2 \right)^{-1} , \\
B & = & \Tilde{H} + a^2\cos^2{\theta}.
\end{eqnarray*}
Multiplying (\ref{spin_metri_loop_1}) to $С(r)$, where  
\begin{equation*}
C(r) = \sqrt{(1 - \frac{M\alpha\beta}{\alpha r^3+\beta M})},    
\end{equation*}
we obtain a rotating solution for the original metrics (\ref{loop}).

\subsection{Conformal Gravity}

Now let's consider gravity with conformal symmetry. These models have a lot of extensions, for example, models with non-linear symmetry realization \cite{Arbuzov:2019rcl, Alexeyev:2020lag}. We apply the metric of BHs in a new massive conformal gravity \cite{Myung_2019,Prokopov:2021lat}:
 \begin{eqnarray}\label{conform} 
G(r) &=& 1 - \frac{2M}{r} + \frac{Q^2_{s}}{r^2} + \frac{Q^2_{s}(Q^2_{s} - M^2 + 6m^{-2}_2)}{3r^4} ,\nonumber \\
F(r) &=& 1 - \frac{2M}{r} + \frac{Q^2_{s}}{r^2} + \frac{2Q^2_{s}(Q^2_{s} - M^2 + 6m^{-2}_2)}{3r^4} ,\nonumber \\
H(r) &=& r^2,
\end{eqnarray}
where $Q_s$ is scalar charge and $m_2$ is massive state with the spin 2.

Applying the modified Newman-Janis method, we obtain a rotating solution for conformal gravity ($\rho^2=r^2 + a^2\cos^2\theta$): 
\begin{eqnarray}\label{spin_metri_conform}
g_{tt} & = & -\frac{\rho^2 + A^2}{\rho^2},\nonumber \\   
g_{t\phi} & = & a\sin^2{\theta}\frac{A^2}{\rho^2},\nonumber \\
g_{\theta \theta} & = & \rho^2,\nonumber \\ 
g_{rr} & = & \frac{\rho^2}{r^2 + a^2  + A^2}, \nonumber \\
g_{\phi \phi} & = & \sin^2{\theta}\left( \rho^2 + a^2\sin^2{\theta}\frac{\rho^2 - A^2}{\rho^2}\right),
\end{eqnarray}
where
\begin{eqnarray*}
A^2 = - 2Mr + Q_s^2 + \frac{2}{3}\frac{Q_s^2}{r^2}\left( Q_s^2-M^2+6m_2^{-2}\right).    
\end{eqnarray*}

\subsection{f(Q) gravity}

We consider the symmetric theory of teleparallelism (STEGR) with a non-zero scalar of non-metricity $Q$ \cite{dambrosio2021black}. A set of spherically symmetric solutions were obtained for this model. The solution that extends GR was chosen ($I^{+}$). Metric can be established as  \cite{dambrosio2021black,Prokopov:2021lat}:
 \begin{eqnarray}\label{f_Q}
G(r) &=& 1 - \frac{2M_{ren}}{r}  + \alpha\frac{32}{r^2}, \nonumber \\
F(r) &=& 1 - \frac{2M_{ren}}{r}  + \alpha\frac{96}{r^2}, \nonumber \\
H(r) &=& r^2,
\end{eqnarray}
where $\alpha$ is the expansion parameter, $c_1$ is integration constant, $M_{ren}$ is renormalized mass. Note that for a distant observer there is no difference between the renormalized and ordinary Schwarzschild masses. So we use $M_{ren}=M$ normalizing all quantities to it.

Applying the modified Newman-Janis method, we obtain a rotating solution for $f(Q)$ gravity ($\rho^2=r^2 + a^2\cos^2\theta$): 
\begin{eqnarray}\label{spin_metri_fQ}
g_{tt} & = & -\frac{\rho^2 + A^2}{\rho^2},\nonumber \\   
g_{t\phi} & = & a\sin^2{\theta}\frac{A^2}{\rho^2},\nonumber \\
g_{\theta \theta} & = & \rho^2,\nonumber \\ 
g_{rr} & = & \frac{\rho^2}{r^2 + a^2  + A^2}, \nonumber \\
g_{\phi \phi} & = & \sin^2{\theta}\left( \rho^2 + a^2\sin^2{\theta}\frac{\rho^2 - A^2}{\rho^2}\right),
\end{eqnarray}
where
\begin{eqnarray*}
A^2 = - 2Mr + 96\alpha.    
\end{eqnarray*}
Rotating solution has the same ansatz as (\ref{spin_metri_conform}). The only difference is in $A(r)$ definitions.

\section{Black Hole Shadows}

\subsection{Modelling Method}

The profile of the black hole's shadow is determined by the last stable orbit for photons so it is necessary to find the solution of the Hamilton-Jacobi equation for isotropic geodesics \cite{Chandrasekhar:1985kt}. As a first step to separate the variables consider the equation where $S$ is the Hamilton-Jacobi function:
\begin{equation}\label{eq:HJE}
    g^{\mu \nu} \frac{\partial S}{\partial x^\mu} \frac{\partial S}{\partial x^\nu} =0. 
\end{equation}
Applying the known integrals of motion $E=-p_t$ и $L_z=p_\phi$ (energy and angular momentum) leads to the following form of the solution:
\begin{equation}
    S = - Et + L_z\phi + S_r(r) + S_\theta(\theta) ,
\end{equation}
After separation of variables:
\begin{eqnarray}
\mathcal{R}(r) & = & \left(K + a^2 - a\lambda\right)^2-(F H +a^2)\left[\eta+\left(a-\lambda\right)^2\right],\nonumber \\
\Theta(\theta) & = & \eta+\cos^2 \theta \left(a^2-\frac{\lambda}{ \sin^2 \theta}\right),
\end{eqnarray}
where $\eta= Q/E^2$, $\lambda=L_z/E$ and $Q$ is Carter's constant. To calculate the last stable orbit it is necessary to find a solution to the equations: 
\begin{equation}
    \mathcal{R} = 0, \qquad \frac{d\mathcal{R}}{dr}=0.
\end{equation}
As a result, one finds the dependence of the quantities $\lambda$ and $\eta$ against metrics. At the last stage a plane normal to the direction on a distant observer is considered. The coordinates of the shadow on such a plane are: 
\begin{eqnarray}
    x' & = & -\frac{\lambda}{\sin\theta_0}, \label{eq45}\\
    y' & = & \pm \sqrt{\eta+a^2 \cos^2\theta_0 - \frac{\lambda^2}{\tan^2\theta_0}}, \label{eq46}
\end{eqnarray}
where $\theta_0$ is solid angle between the plane of rotation and the direction to a distant observer, $\lambda$ и $\eta$ are defined as:
\begin{eqnarray}
    \lambda & = & \frac{K+a^2}{a}-\frac{2K'}{a}\frac{(FH+a^2)}{(HF)'}, \label{eq47} \\
    \eta & = & \frac{4(a^2+FH)}{\left((HF)'\right)^2}\left(K'\right)^2- \nonumber \\ & - & \frac{1}{a^2} \Bigr[ K-\frac{2(FH+a^2)}{(HF)'}K' \Bigr] . \label{eq48}
\end{eqnarray}

To calculate the coordinates $X$ and $Y$ on the picture plane the Eqs  (\ref{eq45}-\ref{eq46}) are used. To solve numerically we improved the Python code written earlier \cite{Prokopov:2021lat,c02}. In addition such quantities as $r_s$ (effective shadow size), $D$ (shadow offset from the centre), $\delta=\Delta_{cs}/r_s$ (shadow distortion during rotation), $\Delta_{cs}$ (the distance between the left border of the shadow and its circular approximation) are sought. Earlier following the same scheme we obtained the shadow profile for the non-local gravity model \cite{c02} to compare with the images of Sgr A* \cite{EventHorizonTelescope:2022apq} and M87*  \cite{EventHorizonTelescope:2023fox}. As before we continue to consider the most probable configurations of Sgr A*: the tilt of the plane of rotation relative to the direction to the observer is $\pi/6$, and the values of the angular acceleration are: $a=0.5$ and $a=0.94$ (relative to the mass $M$) \cite{EventHorizonTelescope:2022apq}. For comparison we show the shadow profile behaviour for the static solution (with $a=0$). The angular acceleration $a=0.9375$ was obtained from observations of the relativistic jet of M87* \cite{Cui:2023uyb}, so the rapid rotation of this black hole is confirmed by observations.

Check the condition of approaching of the circular orbits. We present the function $\mathcal{R}(r)$ and its derivative in the following form (as well as the condition of transition to a circular orbit):
\begin{eqnarray*}
\mathcal{R}(r_{ph}) & = & \mathcal{R}'(r_{ph})=0, \\
\mathcal{R}(r) & = & [\chi(r)E-a L_z]^2-\Delta(r)[K+(L_z-aE)^2], \\
\mathcal{R}'(r) & = & 2E\chi'(r)[\chi(r)E-a L_z]-\Delta'(r)[K+(L_z-aE)^2],
\end{eqnarray*} 
where:
\begin{eqnarray*}
\chi & = & K+a^2, \\
\Delta(r) & = & FH+a^2.
\end{eqnarray*} 
The usage of the definition $\lambda=L_z/E$ leads to $E=L_z/\xi$. This means that the condition for the photon transition to circular orbits can be written as:
\begin{equation*}
\frac{2\chi'}{\Delta'\lambda}=\frac{1}{\Delta}[\frac{\chi}{\lambda}-a]. 
\end{equation*} 
$\lambda$ can be represented as:
\begin{equation*}
\lambda=\frac{\chi}{a}-\frac{2\chi'}{a}\frac{\Delta}{\Delta'}.
\end{equation*} 
Earlier in response to A. F. Zakharov's comment \cite{comment2025} we showed that this expression is satisfied in the case of a non-local metric. Going on we can show that this expression is satisfied for any type of metric by writing the transition condition to circular orbits as:
\begin{equation*}
2\chi\frac{\Delta}{\Delta'}=2\chi\frac{\Delta}{\Delta'}. 
\end{equation*} 
As a result, the expression is reduced and becomes the identity $1=1$. This means that in all cases the condition for the presence of a black hole shadow is satisfied.

\subsection{Horndeski Theory}

To obtain the spinning black hole solution for the Horndeski model (\ref{spin_metri_horn1}) we introduce a new parameter $\alpha = 8\alpha_5\eta/5$. In the first step shadow profiles were obtained for a set of values of $\alpha$ (Fig.~\ref{sh}(a)). In the next step the effective shadow size was estimated (Fig.~\ref{rs}(a)): the shaded area shows the range of values incompatible with the EHT \cite{EventHorizonTelescope:2022apq} data. Further from Fig.~\ref{rs}(a) it is clear that increasing $\alpha$ leads to a decrease in the shadow size. The value $\alpha=1$ is excluded for $a=0.5$ but remains allowed for $a=0.94$. For $\alpha=0.8$, the configuration with $a=0.94$ also remains allowed. For $\alpha<0.5$ both configurations remain in the allowed region. Also note that increasing the value of angular acceleration $a$ leads to an increase in the shadow size i.e. $\alpha$ acts in the direction opposite to $a$ (similar to \cite{c02}). As in the non-rotating case, large values of $\alpha$ are excluded by the EHT results (only $\alpha=1$ is allowed for $a=0.94$). Further, similar to \cite{c02}, the displacement increases linearly with increasing angular acceleration $a$ and does not differ significantly for different values of $\alpha$ (Fig.~\ref{D}(a)). 

The last parameter to be considered is the shadow distortion parameter $\delta$ (Fig.~\ref{delta}(a)). As is easy to see, for all values of $\alpha$ the distortion is approximately $0.5\%-1\%$. In the case of $a=0.94$ the distortion increases (and again $\alpha$ acts in the direction opposite to $a$: with increasing $\alpha$ the distortion decreases) from $2\%$ for $\alpha=1$ to $5.5\%$ in the purely Kerr case.

\subsection{Bumblebee model}

The next metric of a rotating black hole we consider is Eq. (\ref{spin_metri_bb}) for the bumblebee model. As shown earlier \cite{Prokopov:2021lat} in the standard version of this model (where the metric function $G(r)$ has the Schwarzschild value) in the absence of rotation the shadow size corresponds to the Schwarzschild metric ($r_s=3\sqrt{3}M$). Therefore, in Fig.~\ref{sh}(b) we show only the values of $a\neq 0$. As can be seen from Fig.~\ref{rs}(b), for $l \neq 0$ the shadow size becomes smaller than in the Kerr metric. The constraints from Sgr A* permit all values of $l$. Here it is necessary to note that for each value of $l$ a critical value of $a$ exists as was shown in \cite{Capozziello:2023tbo}. This means that for $a=0.94$ only the value of $l$ equal to $l=0.1$ appears to be be correct (for example, for $l=0.2$ the critical value will be $a_{crit}=0.92$ instead of $a=0.94$).

The next important point is that the displacement $D$ is smaller than in the pure Kerr case (Fig.~\ref{D}(b)). This behaviour follows the Horndeski model. The maximum is reached at $a=0.5$. Next, consider the distortion $\delta$ (Fig.~\ref{delta}(b)): at $a=0.5$, the distortion differs from the Kerr value and is approximately $0.8-1.4\%$ depending on $l$. For large values of $a$ the distortion becomes larger than in the pure Kerr case (approximately $5.5\%$) and continues to grow with increasing $l$ (up to $9.2\%$ at $l=0.2$ and $l=0.35$). Finally, for different values of $l$, the distortion values may coincide, since each $l$ has its own value of $a_{crit}$. So if Sgr A* is a rapidly rotating object, the latest results provide an additional opportunity to test and confirm the bumblebee model.

\subsection{Gauss-Bonnet scalar gravity}

The next considered metric of a rotating black hole is Eq. (\ref{spin_metri_gb}) for the Gauss-Bonnet scalar gravity. As noted earlier \cite{Zakharov:2014lqa,Prokopov:2020ewm}, for the model parameter values $\xi>0.3$ the photon sphere ``breaks off'' and the shadow disappears. Therefore only the values $\xi$ lying below this limit are considered. Fig.~\ref{sh}(c) shows the shadow profile. Consider its effective size (Fig.~\ref{rs}(c)): in the static case $a=0$ the shadow decreases with increasing $\xi$. For some non-zero value of $\xi$ the effective size of the shadow grows faster. Further on the size starts to decrease with increasing $\xi$. As in the static case all these combinations are permitted by the EHT results.

Similarly to the bumblebee model the displacement $D$ remains larger than in the Kerr case with $a=0.5$ (Fig.~\ref{D}(c)). The difference between the values of $D$ for the static case and $a=0.5$ is noticeably larger than in the previous case (Fig.~\ref{delta}(c)): for $a=0.5$ the distortion is approximately equal to $0.8-1.2\%$ and remains larger than in the purely Kerr case. For $a=0.94$ a special case occurs: the distortion begins to decrease. Thus, increasing the coupling parameter of the theory $\xi$ causes a decrease in the distortion. For example, for $\xi=0.25$ the distortion is approximately equal to $3.2\%$.

\subsection{Loop Quantum Gravity}

Now we consider the metric of a rotating black hole obtained from the Hayward metric: eq. (\ref{spin_metri_loop_1}). This solution describes a regular black hole without a central singularity. The parameter $l$ characterizes the central energy density $3/8\pi l^2$, $\alpha$ is the time delay between the centre and infinity and $\beta$ is a constant associated with one-loop quantum corrections to the Newtonian potential. Previously \cite{De_Lorenzo_2015,Hu_2018} a constraint on these parameters was obtained: $0\leq \alpha<1$, $\beta_{max}=41/(10\pi)\approx1.305$, and for $l > \sqrt{16/27} M \approx 0.7698$ the object has no horizon. It was also found that in the static case with increasing $l$ the shadow size decreases with increasing $\alpha$ and $\beta$ on the contrary it increases \cite{Prokopov:2021lat}. For $\beta\geq 0$, the minimum shadow size is achieved at $l=\sqrt{16/27}M\approx0.7698$, $\beta=\alpha=0$ equal to $4.92$ and the maximum shadow size occurs at $l=0$, $\beta=41/(10\pi)\approx1.305$, $\alpha=1$ and is equal to $5.32$. At the same time from the constraints of Sgr A*, the shadow size cannot be greater than $5.3$. Earlier \cite{Prokopov:2021lat} it was shown that this configuration can be described by the Reissner-Nordstrom space-time and therefore, using only the shadow size, it is impossible to obtain the values of all parameters without additional observational data. Therefore, when modelling the rotating case, special attention was paid to the boundary configurations.

Shadow profiles were obtained for a wide range of additional parameters (Fig.~\ref{sh}(d)) and the effective shadow radius was estimated for each case (Fig.~\ref{rs}(d)). As can be seen from the last figure applying the analogy with the bumblebee model for each value of $l$ there is a critical value $a$. Moreover the larger $l$ the smaller $a_{crit}$ (for example, for the maximum value $l=\sqrt{16/27}\approx0.7698$ the value $a_{crit}=0.66$). However increasing $\alpha$ or $\beta$ leads to an increase of $a_{crit}$ (for example, in the case of $l=\alpha=0.5$ and $\beta=0.2$ this value will be $a_{crit}=0.86$. With the same values but $\beta=0.8$ it is possible to simulate the maximum value of the rotation parameter in the available configurations $a=0.94$). Further, from (Fig.~\ref{rs}(d)) it can be concluded that this tendency is the same as in the static case. The effective size of the shadow will be larger than in the Kerr case only at the maximum value of the parameters $\alpha$, $\beta$ and $l=0$ (this case does not contradict the restrictions from Sgr A* if $a\neq0$). All other configurations are possible. Next, from the graph of the dependence of the displacement parameter $D$ on the angular acceleration (Fig.~\ref{D}(d)) it is clear that $D$ increases linearly with rotation in the case of $l=0$, but for $l\neq0$ the growth becomes non-linear near the critical value of the rotation parameter $a_{crit}$.

Move on to the last parameter: the distortion $\delta$ (Fig.~\ref{delta}(d)). As follows from the graph for $a=0.5$ the distortion is approximately $0.5\%-1\%$ in all cases except for $l_{max}=0.7698$ (where it is equal to $3\%$). The case $a=0.94$ is possible for $l=0$ and maximum of $\alpha$ and $\beta$: the distortion will be less than in the Kerr case and equal to $3.4\%$ (in Kerr it is $5.5\%$). For $l=\alpha=0.5$, $\beta=0.8$ the distortion during fast rotation will exceed the Kerr distortion, reaching $13\%$. For $l_{max}=0.7698$ with $a=a_{crit}=0.66$ the distortion is $12\%$. In the last case $l=\alpha=0.5$, $\beta=0.2$ with $a=a_{crit}=0.86$ the distortion will be $11\%$.

\subsection{Conformal gravity}

Consider the metric of a rotating black hole in massive conformal gravity: eq. (\ref{spin_metri_conform}) It was previously shown \cite{Prokopov:2021lat} that large $Q_s$ and $1/m_2$ lead to the absence of a photon sphere. Decreasing the value of $m_2$ leads to a decrease in the size of the shadow. The EHT constraints exclude only large if $Q_s$ and $m_2$ (for example, if $m_2=2$, then $Q_s<0.9$).

As a first step, shadow profiles were constructed for a wide range of the additional parameters $Q_s$ and $m_2$ (Fig.~\ref{sh}(e)). The second step was to calculate the effective radius of the shadow (Fig.~\ref{rs}(e)). As can be seen from the last graph from the proposed configurations the state with $m_2=\infty$ and $Q_s=0.9$ is prohibited by the Sgr A* constraints. As $m_2$ increases the shadow size increases as in the static case. But an increase in $Q_s$ on the contrary decreases the shadow size. Also, for many configurations  a critical value of the rotation parameter $a_{crit}$ exists. For $m_2=2$ and $Q_s=0.73$ it is $a_{crit}=0.5$, for $m_2=2$ and $Q_s=0.5$ --- $a_{crit}=0.82$, for $m_2=\infty$ and $Q_s=0.9$ --- $a_{crit}=0.55$, for $m_2=\infty$ and $Q_s=0.5$ --- $a_{crit}=0.93$, for $m_2=1$ and $Q_s=0.51$ --- $a_{crit}=0.5$, for $m_2=1$ and $Q_s=0.35$ --- $a_{crit}=0.78$, for $m_2=1$ and $Q_s=0.2$ --- $a_{crit}=0.93$, for $m_2=0.707$ and $Q_s=0.39$ --- $a_{crit}=0.5$, for $m_2=0.707$ and $Q_s=0.2$ --- $a_{crit}=0.87$. Consequently of all the proposed configurations slow rotation of Sgr A* ($a=0.5$) is always possible while fast rotation ($a=0.94$) is practically everywhere prohibited (although in some cases the critical rotation parameter is close to this value). In the third step, we plot the dependence of the displacement $D$ on rotation (Fig.~\ref{D}(e)). As follows from this graph, the displacement near $a=0.5$ is greater than when close to $a=0.94$.

At the last step we plot the dependence of the $\delta$ distortion on the rotation (Fig.~\ref{delta}(e)). As can be seen from the graph, the distortions at $a=0.5$ when $a_{crit}$ is significantly greater than $0.5$ correspond to the Kerr distortions and are equal to $0.5\%-1.5\%$. When $a_{crit}$ is close to $0.5$, the distortion becomes equal to $3\%-4.5\%$. In this case, the maximum distortion is realized at $m_2=\infty$ and $Q_s=0.9$ and at $a_{crit}=0.55$. With the same parameters and $a=0.5$, the distortion is $7\%$, which is already greater than the Kerr distortion at $a=0.94$ ($5.5\%$). It is also worth noting the strong deviations of the Kerr black hole with rapid rotation. For $a=0.78$, the configurations $m_2=2$ and $Q_s=0.5$, as well as $m_2=1$ and $Q_s=0.35$ have approximately the same distortion as the Kerr black hole for $a=0.94$ ($5.5\%$). The distortion is $7.5\%$ for $a=0.82$, $m_2=2$ and $Q_s=0.5$, as well as for $a=0.93$, $m_2=1$ and $Q_s=0.2$. The largest distortion is achieved for $a=0.93$, $m_2=\infty$, $Q_s=0.5$ and is $9.5\%$.

\subsection{$f(Q)$ gravity}

\begin{figure}[htbp]
\centering
\includegraphics[width=\textwidth]{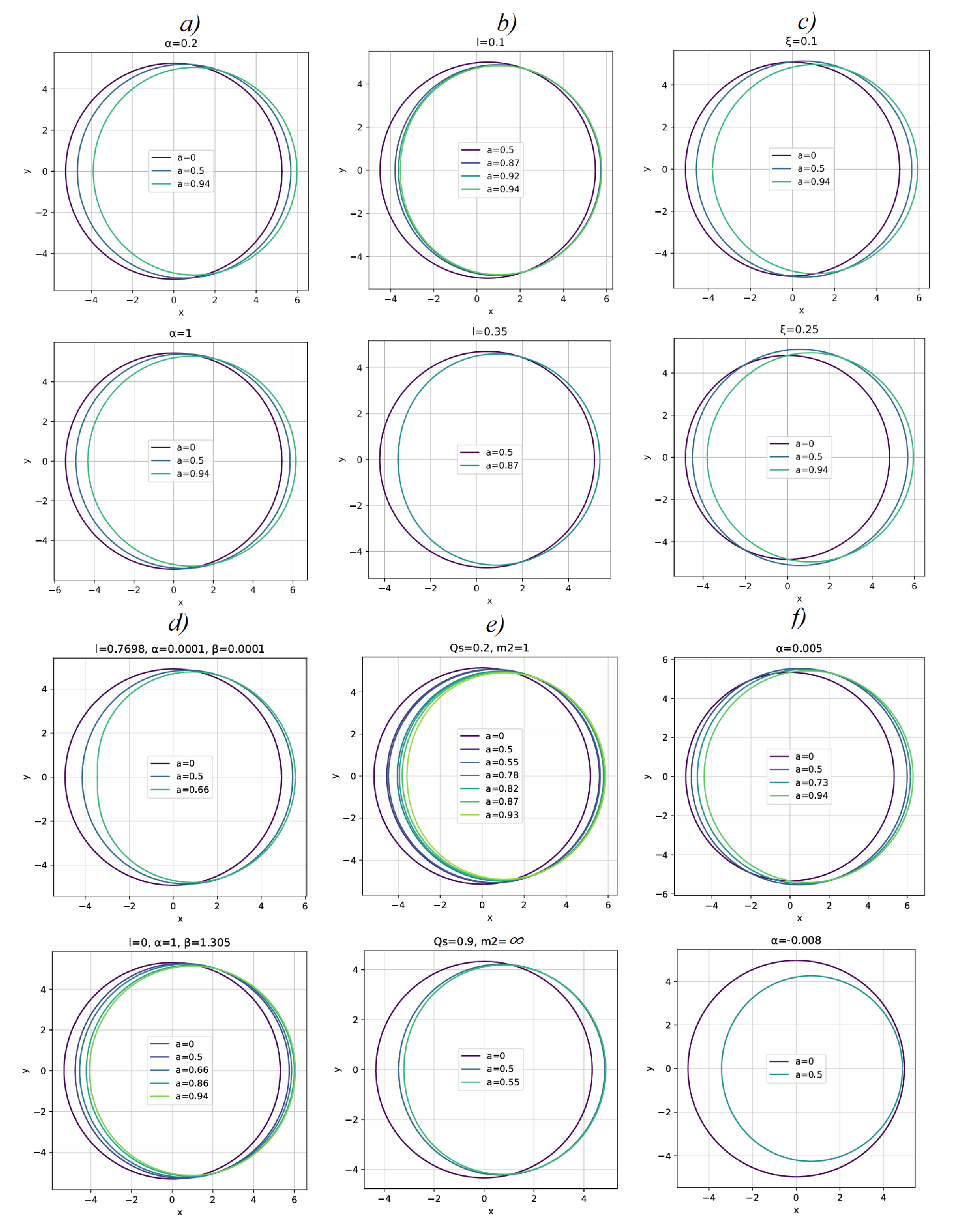}
\caption{Black hole shadow profiles against angular acceleration $a$ for different values of (a) $\alpha$ in the Horndeski model, (b) $l$ in the Bumblebee model, (c) $\xi$ in scalar Gauss-Bonnet gravity, (d) $l$, $\alpha$, $\beta$ in loop quantum gravity, (e) $m_2$ and $Q_s$ in conformal gravity, and (f) $\alpha$ in $f(Q)$ gravity. The minimum and maximum of the additional parameters of the theories are shown. The tilt angle of the plane of rotation is $\theta_0=\frac{\pi}{6}$ (Sgr A*).}
\label{sh}
\end{figure}

\begin{figure}[htbp]
\centering
\includegraphics[width=\textwidth]{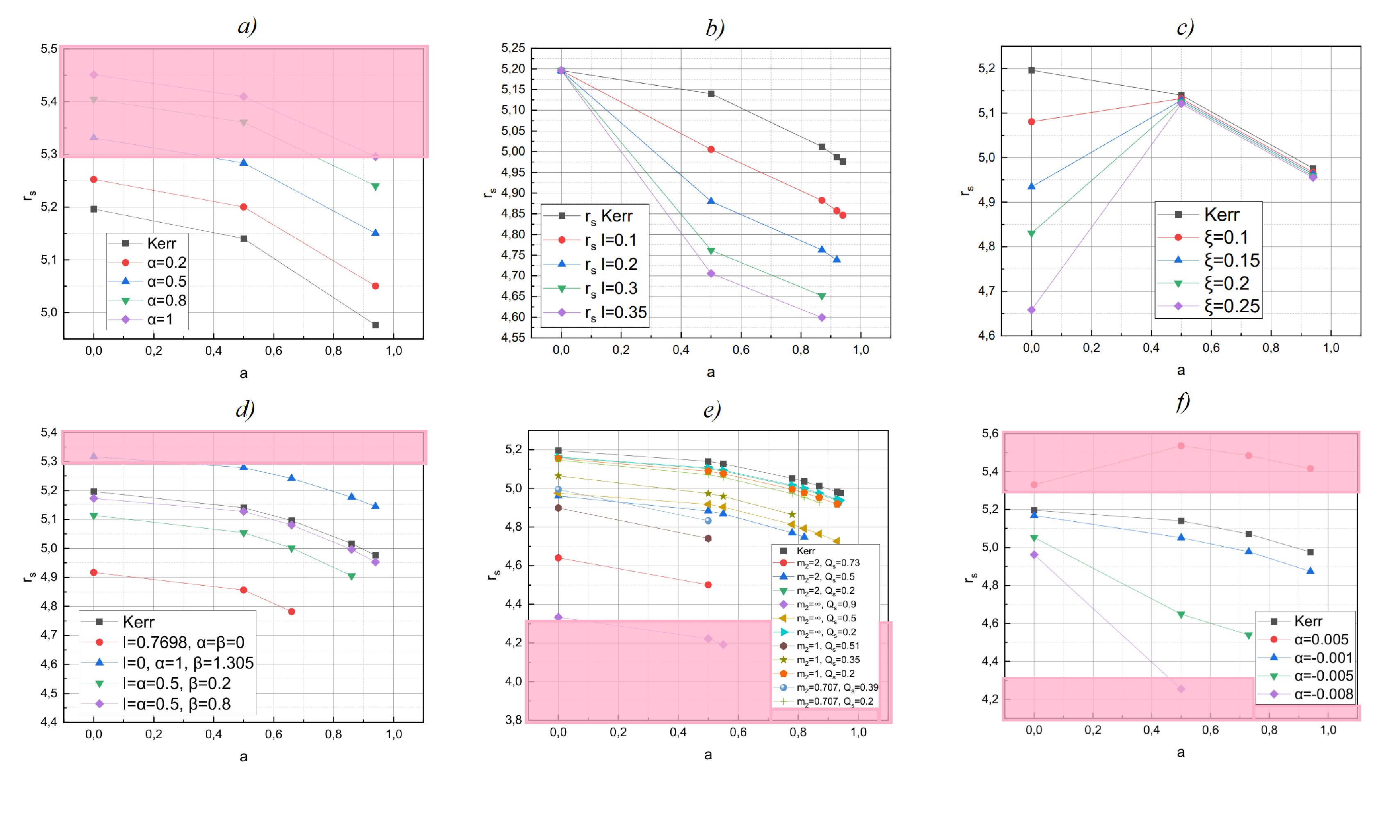}
\caption{Shadow size $r_s$ versus angular acceleration $a$ for different values of (a) $\alpha$ in the Horndeski model, (b) $l$ in the Bumblebee model, (c) $\xi$ in scalar Gauss-Bonnet gravity, (d) $l$, $\alpha$, $\beta$ in loop quantum gravity, (e) $m_2$ and $Q_s$ in conformal gravity, and (f) $\alpha$ in $f(Q)$ gravity. The minimum and maximum of the additional parameters of the theories are shown. The tilt angle of the plane of rotation is $\theta_0=\frac{\pi}{6}$ (Sgr A*). The red region is forbidden by the Sgr A* results.}
\label{rs}
\end{figure}

\begin{figure}[htbp]
\centering
\includegraphics[width=\textwidth]{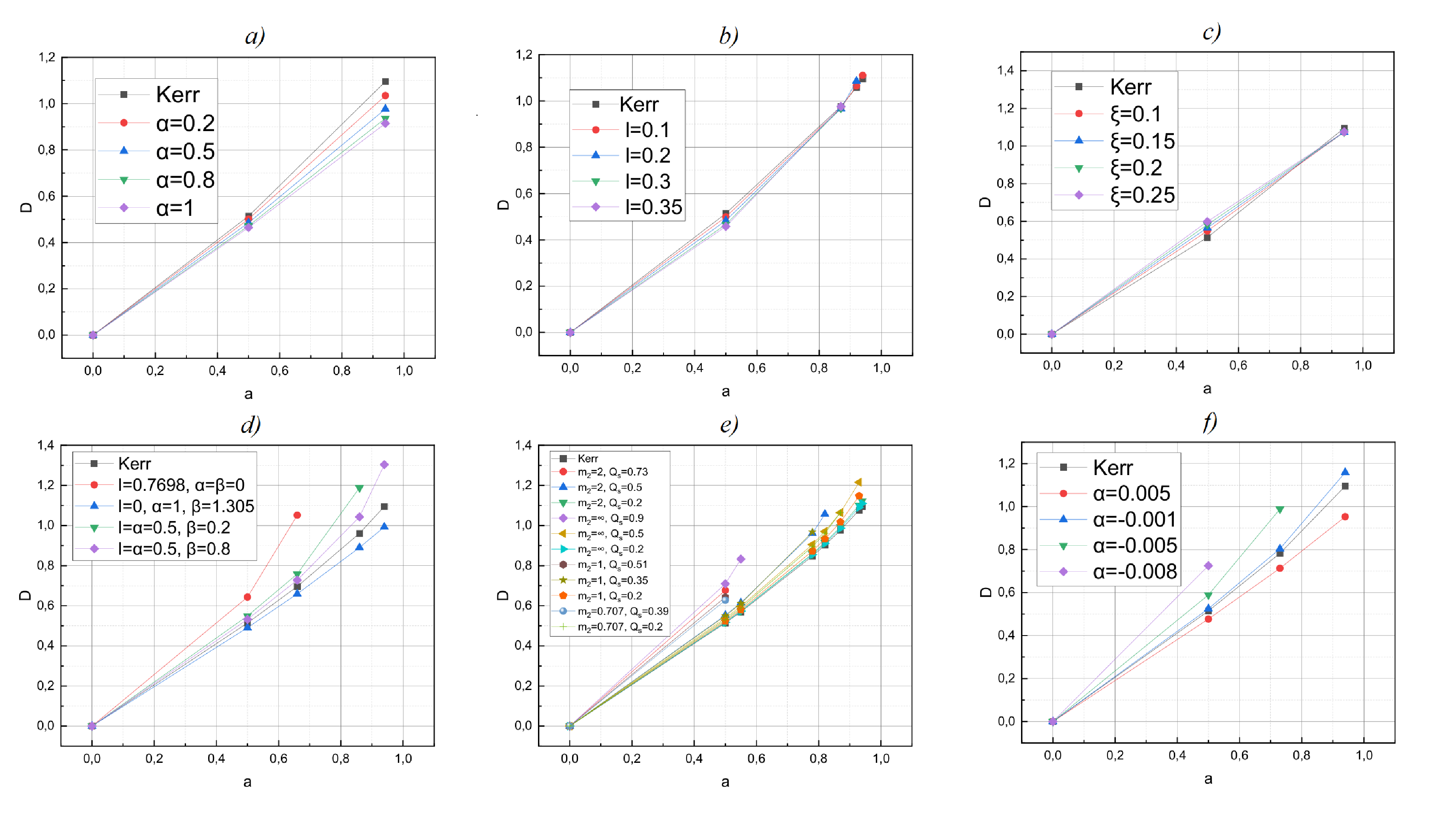}
\caption{Dependence of the displacement $D$ versus the angular acceleration $a$ for different values of (a) $\alpha$ in the Horndeski model, (b) $l$ in the Bumblebee model, (c) $\xi$ in scalar Gauss-Bonnet gravity, (d) $l$, $\alpha$, $\beta$ in loop quantum gravity, (e) $m_2$ and $Q_s$ in conformal gravity, and (f) $\alpha$ in $f(Q)$ gravity. The minimum and maximum of the additional parameters of the theories are shown. The tilt angle of the plane of rotation is $\theta_0=\frac{\pi}{6}$ (Sgr A*).}
\label{D}
\end{figure}

Earlier \cite{Prokopov:2021lat} it was shown that the constraint on the parameter $\alpha$ in the static case is as follows: $-0.025<\alpha<0.005$. That is the average value of $\alpha$ shifts closer to the negative region. If we believe to this trend with further refinement of the observational data $\alpha$ will shift to the negative region. Taking this into account, we construct the profiles of the black hole shadow for various $\alpha$ (Fig.~\ref{sh}(f)) and the effective radius of the shadow (Fig.~\ref{rs}(e)). We note that the value $\alpha=0.005$ does not pass from the upper constraints on the size of the black hole shadow from Sgr A* in both rotating cases (the static case is on the boundary). The case $\alpha=-0.008$ does not pass from the lower constraints on the shadow size of the EHT. It is also worth noting the reason why it is not possible to construct rotating cases for large absolute values of $\alpha$ in the negative region. For positive $\alpha$ simulation with fast rotation is possible. However in the case of $\alpha<0$, for each value of $\alpha$ there is a critical value of the rotation parameter $a_{crit}$. For $\alpha=-0.001$, this is $a_{crit}=0.94$, for $\alpha=-0.005$ --- $a_{crit}=0.73$, and for $\alpha=-0.008$ --- $a_{crit}=0.5$. After constructing the dependence of the displacement $D$ on the rotation $a$ (Fig.~\ref{D}(f)) one can see that in the case of positive $\alpha$ the displacement will be smaller than in the case of a Kerr black hole and for $\alpha<0$ it will be larger especially near $a_{crit}$.

\begin{figure}[htbp]
\centering
\includegraphics[width=\textwidth]{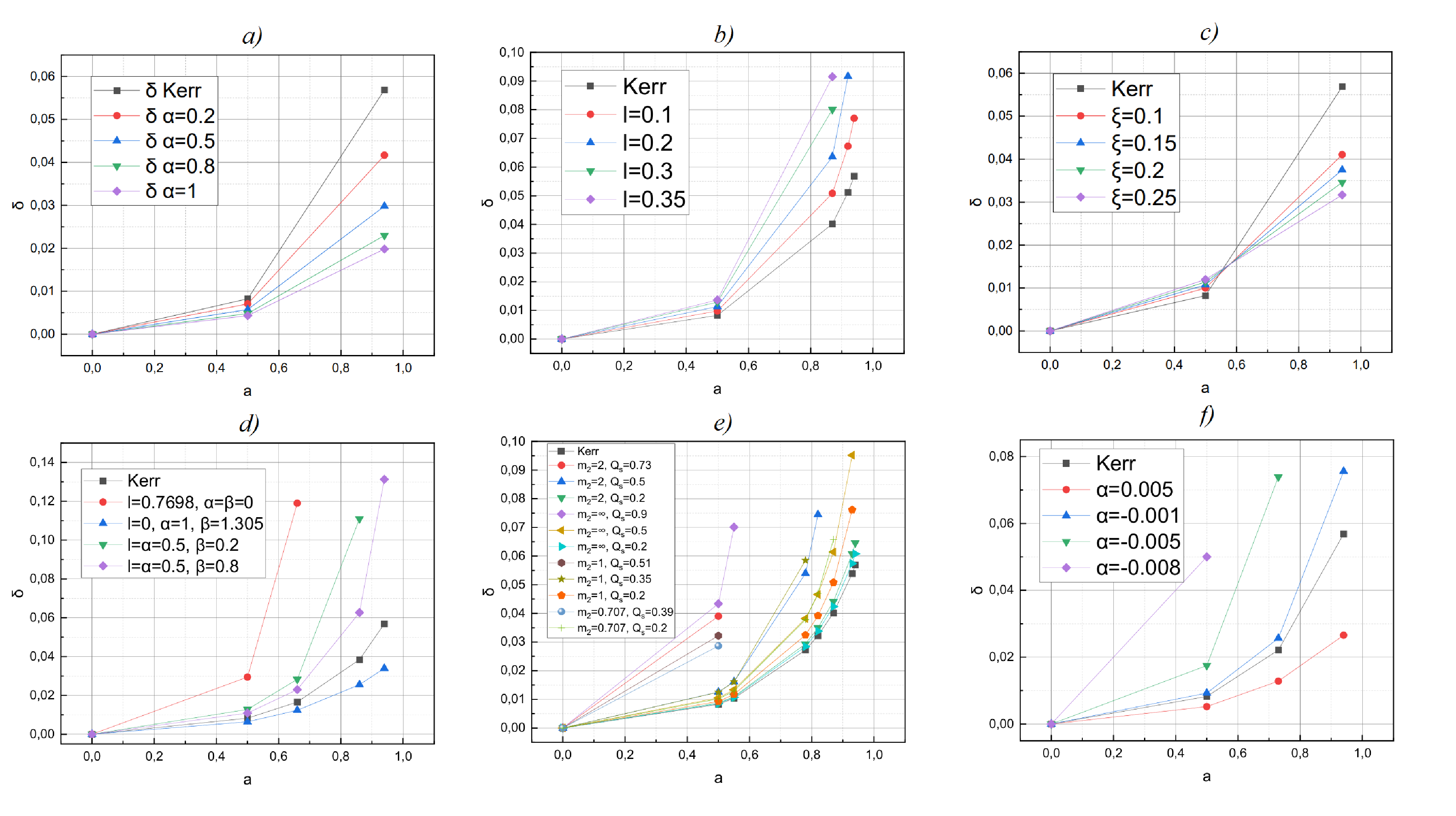}
\caption{Dependence of the distortion $\delta$ upon the angular acceleration $a$ for different values of (a) $\alpha$ in the Horndeski model, (b) $l$ in the bumblebee model, (c) $\xi$ in scalar Gauss-Bonnet gravity, (d) $l$, $\alpha$, $\beta$ in loop quantum gravity, (e) $m_2$ and $Q_s$ in conformal gravity, and (f) $\alpha$ in $f(Q)$ gravity. The minimum and maximum of the additional parameters of the theories are shown. The tilt angle of the plane of rotation is $\theta_0=\frac{\pi}{6}$ (Sgr A*).}
\label{delta}
\end{figure}

Next, we plot the dependence of the $\delta$ distortion on the $a$ rotation (Fig.~\ref{delta}(f)). For $\alpha>0$, the distortion will be less than in the Kerr case, and for $\alpha<0$, on the contrary, it will be greater. In the case of slow rotation, the distortion for $\alpha=0.005$ is approximately equal to $0.5\%$, for $\alpha=-0.001$ --- to the Kerr $1\%$, and for $\alpha=-0.005$ --- it is greater and approximately equal to $1.7\%$. The greatest distortion for $a=0.5$ is realized in the case of $\alpha=-0.008$ for which this value of the rotation parameter is critical and the distortion is equal to $5\%$. For the case of fast rotation it is seen that for $\alpha=-0.005$ and $\alpha=-0.001$ the distortion is approximately equal to $7.5\%$ for their critical rotation parameters ($a_{crit}=0.73$ and $a_{crit}=0.94$ respectively), which is greater than the Kerr distortion for $a=0.94$ ($5.5\%$).

\onecolumn

\twocolumn

\section{Discussion and conclusions}

Applying the improved version of the Newman-Janis algorithm rotating solutions are obtained for a set of models that extend general relativity in different ways. This set includes the Horndeski model (the most general case of scalar-tensor gravity with second-order field equations), bumblebee (gravity with a vector field), scalar Gauss-Bonnet gravity, loop quantum gravity, conformal gravity, and $f(Q)$ being a special case of symmetric teleparallel gravity STEGR, for whose static versions the constraints have already been obtained \cite{Prokopov:2021lat}. Since both black holes photographed by the EHT project: Sgr A* and M87* are rotating \cite{EventHorizonTelescope:2022apq,EventHorizonTelescope:2023fox} the next step: using rotating metrics for theoretical modelling of the shadows of these black holes --- seems justified. Since for Sgr A* the most probable values of the angular momentum are $a=0.5$ and $a=0.94$, and for M87* it is $a=0.9375$ these are the values we used in the modelling.

When analysing the shadow profiles and all the characteristics together, we can draw the following conclusions:

\begin{enumerate}

\item Spherically symmetric solutions for extended gravity theories contain a number of additional parameters that are not present in the simplest solution of GR --- the Schwarzschild metric. Further, these solutions, in addition to having one or more additional parameters, have a more complex structure compared to the Reissner-Nordstrom metric. Therefore, the resulting metrics of rotating black holes have a structure that is more complex compared to the Kerr-Newman metric. Further effects follow from this.

\item The presence of additional parameters of the theory due to the more complex structure of the solution gives rise to the presence of critical values of the angular momentum $a_{crit}$. Such values exist in all the theories considered, except for the Horndeski model and, partially, the Gauss-Bonnet scalar-tensor gravity (and even then, it is necessary to consider the values $\xi<0.3$, at which the existence of the photon sphere is ensured).

\item As a result of a comprehensive consideration of the spectrum of theories, the previously made conclusion is confirmed \cite{c02,c03} that for some of the models considered, taking into account the parameters of the theory either slows down the rotation and the effects associated with it (this is most clearly manifested for the Horndeski theory and the Gauss-Bonnet scalar gravity), or enhances them (this is most clearly manifested for the bumblebee model). For the other models considered, this effect is also present, but it does not work so linearly. Thus, taking into account the results in non-local gravity, we can conclude that the extended theory of gravity corrects the effects of rotation in both directions. This is important for further modelling of shadow profiles, taking into account the constantly increasing accuracy of photographing black holes.

\item Considering the dependence of the displacement parameter and its closeness to the Kerr value, we can conclude that the first metrics of rotating black holes for the three theories considered: the Horndeski, bumblebee, and Gauss-Bonnet scalar gravity models, work best and with a minimum number of additional parameters and restrictions as a basis for modelling the shadow profiles of black holes. Apparently, the best results should be expected from the Horndeski model (taking into account that new types of solutions are possible in this theory, since so far all the solutions considered in the literature represent special cases of the theory). The bumblebee model provides the best agreement with the Kerr metric.

\item Despite the less accurate modelling of shadow profiles than the first three metrics, we note that the Hayward metric: the black hole without a central singularity, is of additional interest, since loop quantum gravity apparently manages to get rid of both curvature singularities: at the centre of the black hole (the presented Hayward metric) and at the beginning of cosmological evolution, replacing the singularity with a bounce and ensuring the existence of the inflationary stage \cite{barrau}.

\end{enumerate}

Thus, taking into account the rotation, photographs of black hole shadows, along with the GW170817 test or the post-Newtonian formalism \cite{Alexeyev:2022mqb}, can already serve as a way to test and constrain extended theories of gravity.   

\section{Acknowledgements}

The investigation is supposed by the Russian Science Foundation via grant RSF 23-22-00073.

\end{document}